\documentclass[12pt]{article}
\usepackage{graphicx}
\usepackage{epsfig}
\usepackage[utf8x]{inputenc}
\usepackage[T2A]{fontenc}
\usepackage[english]{babel}
\usepackage{textcomp}
\usepackage{mathtools}
\usepackage{dcolumn}
\usepackage{epsfig}
\usepackage{epstopdf}
\usepackage{bm}
\usepackage[left=1cm,right=1cm,top=2cm,bottom=2cm]{geometry}
\usepackage[usenames]{color}
\usepackage{colortbl}

\newcommand{\be}{\begin{equation}}
\newcommand{\ee}{\end{equation}}

\newcommand{\bea}{\begin{eqnarray}}
\newcommand{\eea}{\end{eqnarray}}
\newcommand{\beq}{\begin{equation}}
\newcommand{\eeq}{\end{equation}}

\def\fun#1#2{\lower3.6pt\vbox{\baselineskip0pt\lineskip.9pt
		\ialign{$\mathsurround=0pt#1\hfil##\hfil$\crcr#2\crcr\sim\crcr}}}

\begin{document}
	\title{Tidal Forces in Kottler Spacetimes}
	\date{}
	\author{V.P. Vandeev$^*$, A.N. Semenova$^+$,}
	\maketitle
	\begin{center}
		{\it $^*$BALTIC STATE TECHNICAL UNIVERSITY «VOENMEH» named after D.F. Ustinov, 1-Ya Krasnoarmeyskaya Ulitsa, 1, St Petersburg, 190005, Russia}
		
		{\it $^+$Petersburg Nuclear Physics Institute of National Research Centre ``Kurchatov Institute'', Gatchina, 188300, Russia}
	\end{center}
	
	\begin{abstract}
	The article considers tidal forces in the vicinity of the Kottler black hole. We find a solution of the geodesic deviation equation for radially falling bodies, which is determined by elliptic integrals. And also the asymptotic behavior of all spatial geodesic deviation vector components were found. We demonstrate that the radial component of the tidal force changes sign outside the single event horizon for any negative values of the cosmological constant, in contrast to the Schwarzschild black hole, where all the components of the tidal force are sign-constant. We also find the similarity between the Kottler black hole and the Reissner -- Nordstr{\"{o}}m black hole, because we indicate the value of the cosmological constant, which ensures the existence of two horizons of the black hole, between which the angular components of the tidal force change sign. It was possible to detect non-analytical behavior of geodesic deviation vector components in Anti-de Sitter spacetime and to describe it locally.
\end{abstract}

\section{\label{sec:level1}Introduction}
The physical properties of black holes are still very interesting for the scientific community, despite the fact that for the first time a black hole, as a solution to Einstein's equations, was discovered in GR more than a hundred years ago. That was the Schwarzschild solution \cite{MS}, which has a single parameter - the mass of the black hole. Subsequently, many solutions of Einstein's equations appeared, which generalized Schwarzschild's solution: Kerr (rotating) black hole \cite{MK}, Reissner -- Nordstr{\"{o}}m (electrically charged) black hole \cite{MR}, \cite{MN}, Kerr -- Newman (charged and rotating) black hole \cite{MKN}. However, non-vacuum solutions are of particular interest. And the simplest solution of this type is the Kottler metric \cite{MKot}, which solves the equations of general relativity when the cosmological constant plays the role of matter. It is also called the Schwarzschild -- De Sitter metric, and its physical \cite{AP} and geometric \cite{AG} properties are actively explored.

In this article, we focus on the studying of tidal forces in the Kottler metric, like articles \cite{TFrn} and \cite{TFkbh}, which discuss the same issues only for Reisnner -- Nordstr{\"{o}}m and Kiselev black holes, respectively.  General spherical symmetry spaces are considered in \cite{SSSS}. The tidal effect in spaces with unusual regular black hole metrics and Schwarzschild metric with massive graviton is considered in \cite{Terbh} and \cite{MG}.
It is well known that in Schwarzschild spacetime a test body falling towards the event horizon of BH experiences stretching in the radial direction and compression in the both angular directions \cite{GR}. However, the presence of matter, which is represented by a cosmological constant, significantly changes the behavior of tidal forces in comparison with empty space. The most important differences are that at infinity all the components of the tidal forces are nonzero and also they cease to be a constant sign. At certain points of Kottler metric, the tidal forces in the radial or angular direction change their sign unlike in Schwarzschild metric. The paper is organized as follows. In Sec.~2 we discuss properties of Kottler spacetime. We analyze geodesics in Kottler metric in Sec.~3. And study tidal forces of such black holes in Sec.~4. In Sec.~5 we obtain the solutions of the geodesic deviation equations. Then in Sec.~6. we provide an asymptotic analysis of solutions in the vicinity of a physical singularity and infinitely far from it. And Sec.~7 includes an analysis of geodesic deviation equations solutions in the Schwarzschild -- Anti-de Sitter spacetime, there we manage to show the presence of non-analytical behavior of all deviation vector components at large distances from the black hole horizon. We use the metric signature $(+,-,-,-)$ and set the speed of light $c$ and Newtonian gravitational constant $G$ to $1$ throughout this paper.

\section{\label{sec:level2}Kottler Black Holes}
In this paper, we investigate the properties of a static solution of the Einstein equation in the presence of cosmological constant
\begin{equation}\label{EEQ}
    R_{\mu\nu}-\frac{R}{2}g_{\mu\nu}+{\Lambda}g_{\mu\nu}=0.
\end{equation}
The line element of a static black hole in the presence of constant energy density is given by \cite{MKot}
\begin{equation}\label{interval}
ds^2=g_{\mu\nu}dx^{\mu}dx^{\nu}=f(r)dt^2-\frac{dr^2}{f(r)}-r^2\left(d\theta^2+\sin^2{\theta}d\varphi^2\right),
\end{equation}
with
\begin{equation}\label{f}
f(r)=1-\frac{2M}{r}-\frac{{\Lambda}r^2}{3},
\end{equation}
where $M$ is mass of a black hole and $\Lambda$ is cosmological constant.
To find the radius of the event horizon of a black hole, we find solutions of the equation $f(r)=0$. Passing to the dimensionless parameters $\rho=\frac{r}{M}$ and $\lambda={\Lambda}M^2$ horizon equation takes the form
\begin{equation}\label{horeq}
f(\rho)=\frac{-{\lambda}\rho^3+3\rho-6}{3\rho}=0.
\end{equation}

There are four possible cases of the structure of the Kottler spacetime. They are described in \cite{GH} and come down to:

\begin{enumerate}
  \item For ${\Lambda}M^2\leq0$ we have single event horizon $R_{\Lambda}$, which matches with the Schwarzschild horizon $R_s=2M$ at ${\Lambda=0}$.
  \item For $0<{\Lambda}M^2<\frac{1}{9}$ there are two horizons of the black hole, which we call $R_{+}$ and $R_{-}$.
  \item For ${\Lambda}M^2=\frac{1}{9}$ there is only one critical horizon $R_c=3M$.
  \item For ${\Lambda}M^2>\frac{1}{9}$ black hole does not have any horizons.
\end{enumerate}

Using the Cardano formula \cite{LA} we obtain dimensionless positive roots of (\ref{horeq})
\begin{equation}\label{lh}
\rho_{\Lambda}=\frac{R_\Lambda}{M}=\sqrt[3]{-\frac{3}{\lambda}+\frac{1}{\lambda}\sqrt{9-\frac{1}{\lambda}}}+
\sqrt[3]{-\frac{3}{\lambda}-\frac{1}{\lambda}\sqrt{9-\frac{1}{\lambda}}},\\
\end{equation}
\begin{equation}
\left[
      \begin{gathered}
    \rho_{-}=\frac{R_-}{M}=\frac{2}{\sqrt{\lambda}}\cos{\left(\frac{\arccos{\left(-3\sqrt{\lambda}\right)}+4\pi}{3}\right)},\\
    \rho_{+}=\frac{R_+}{M}=\frac{2}{\sqrt{\lambda}}\cos{\left(\frac{\arccos{\left(-3\sqrt{\lambda}\right)}}{3}\right)},
  \end{gathered}
\right.
\end{equation}

\begin{equation}\label{ch}
\rho_c=\frac{R_c}{M}=3.
\end{equation}

\begin{figure}[h!]
\center{\includegraphics[width = 9.5 cm]{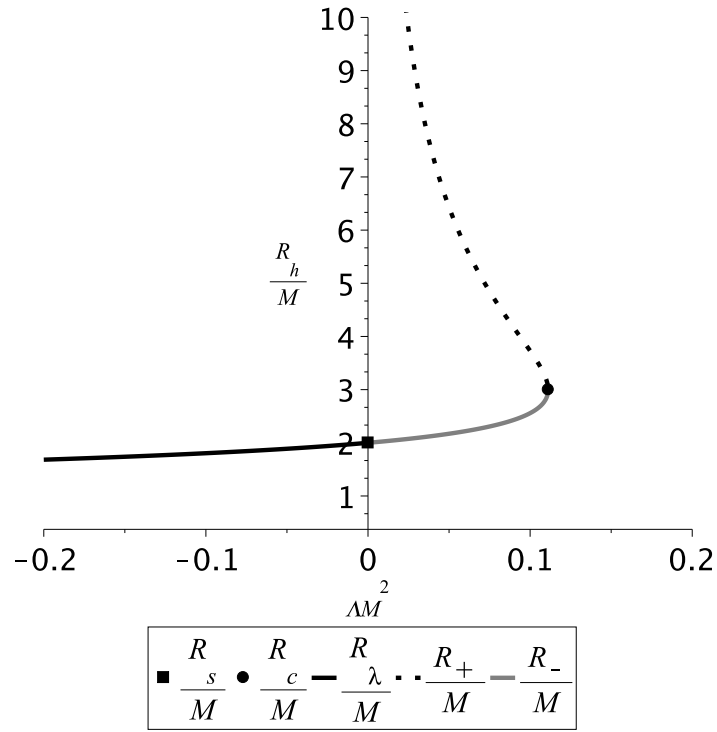}
\caption{\footnotesize{Event horizon radius $\rho_h=\frac{R_h}{M}$ dependence on cosmological constant $\lambda=\Lambda M^2$.}}
\label{hor}}
\end{figure}

As a result, we have the opportunity to build a dependence of the radius of the event horizon of a black hole on the magnitude of the cosmological constant, the value of which is presented in Fig. \ref{hor}. Where the square and circle indicate $R_s$ and $R_c$ correspondingly.

Further in the article, we will call an arbitrary black hole horizon $R_h$ (or its dimensionless analog $\rho_h=\frac{R_h}{M}$) if we are not interested in the specific properties of the horizon or horizons.

\section{\label{sec:level3}Radial geodesics in Kottler Spacetime}
In radial motion, there are two meaningful geodesic equations \cite{mtbh} for the time and radial variables
\begin{equation}\label{teq}
\frac{dt}{d\tau}=\frac{E}{f(r)},
\end{equation}
\begin{equation}\label{req}
\left(\frac{dr}{d\tau}\right)^2=E^2-f(r).
\end{equation}
where $d\tau$ is the proper time. And it should be noted that the dynamics in angular variables is trivial because we study radial geodesics and therefore we set the angular momentum to zero. For the radial infall of a test particle from rest at position $b$, we obtain $E^2=f(b)$ from $\frac{dr}{d\tau}=0$ \cite{st}.

By defining the ''Newtonian radial acceleration'' \cite{mech} as $A^r\equiv\ddot{r}$, we find from Eq. (\ref{req}) that
\begin{equation}
    A^r=-\frac{f'(r)}{2},
\end{equation}
where the prime denotes the differentiation with respect to the radial coordinate $r$. For Kottler metric this becomes
\begin{equation}
    A^r=-\frac{M}{r^2}+\frac{{\Lambda}r}{3}.
\end{equation}
Where it can be seen that a negative value of $\Lambda$ gives us additional attraction to a black hole, and for positive $\Lambda$, regions of attraction and repulsion appear. These areas are separated by the surface of a sphere of radius $r_{eq}$ which is the solution of equation $A^r(r_{eq})=0$
\begin{equation}
    r_{eq}=\sqrt[3]{\frac{3M}{\Lambda}}.
\end{equation}
It is interesting to note that the test particle falling freely from rest at $r=b>R_h$ can turn back at $R^{stop}$. The radius $R^{stop}$ can easily be found as a root of $E^2=f(r)$ as
\begin{equation}\label{rstop}
R^{stop}=-\frac{b}{2}+\sqrt{\frac{b^2}{4}+\frac{6M}{{\Lambda}b}},
\end{equation}
which is real and positive when ${\Lambda}>0$. It should be noted that for $\Lambda = 0$ (Schwarzschild metric) $R_{stop}$ is absent. Eq.~(\ref{rstop}) allows us to compare points $R_{stop}$ and $b$:
\begin{equation}
\left[
\begin{aligned}
&R^{stop}\geq b\: \text{at}\: \Lambda\in \bigg(0,\frac{3M}{b^3}\bigg],\\
&R^{stop}<b\:\text{at}\:\Lambda\in \bigg(\frac{3M}{b^3},+\infty\bigg).
\end{aligned}
\right.
\end{equation}
We are interested in case where $R_{stop}<b$ because a turn of test particle is possible in $R_{stop}$ point.
\section{\label{sec:level4}Tidal forces in Kottler Spacetime}
\subsection{Geodesic deviation equation}
Now let us consider tidal forces in the Kottler metric. As is well known \cite{GR}, the equation for the spacelike components of the geodesic deviation vector $\tilde{\xi}^{\mu}$ that describes the distance between two infinitesimally close particles in free fall is given by
\begin{equation}\label{deq}
\frac{D^2\tilde{\xi}^{\mu}}{d\tau^2}=R^{\mu}_{\nu\alpha\beta}u^{\nu}u^{\alpha}\tilde{\xi}^{\beta},
\end{equation}
where $u^{\nu}$ is the unit vector of 4-velocity tangent to the geodesic. The tetrad basis for radial free-fall reference frames have form:
\begin{subequations}
\begin{equation}
e_{t}^{\mu}=\left(\frac{E}{f},\sqrt{E^2-f},0,0\right),
\end{equation}
\begin{equation}
e_{r}^{\mu}=\left(-\frac{\sqrt{E^2-f}}{f},-E,0,0\right),
\end{equation}
\begin{equation}
e_{\theta}^{\mu}=\left(0,0,\frac{1}{r},0\right),
\end{equation}
\begin{equation}
e_{\varphi}^{\mu}=\left(0,0,0,\frac{1}{r\sin{\theta}}\right).\\
\end{equation}
\end{subequations}
where $e^{\mu}_{\alpha}$ satisfy normalization condition $e^{\mu}_{\alpha}\:e^{\nu}_{\beta}\:g_{\mu\nu}=\eta_{\alpha\beta}$ and $\eta_{\alpha\beta}$ is Minkowski metric. The geodesic deviation vector can be submitted as
\begin{equation}\label{lt}
    \tilde{\xi}^{\mu}=e^{\mu}_{\nu}\:\xi^{\nu}.
\end{equation}
The nonzero components of the Riemann tensor are calculated by the Kottler metric from expression (\ref{interval}) and have the form
\begin{subequations}
\begin{equation}
R^{0}_{.101}=-\frac{f''}{2f},\\
\end{equation}
\begin{equation}
R^{0}_{.202}=R^{1}_{.212}=-\frac{f'r}{2},\\
\end{equation}
\begin{equation}
R^{0}_{.303}=R^{1}_{.313}=-\frac{f'r\sin^2{\theta}}{2},\\
\end{equation}
\begin{equation}
R^{2}_{.323}=\left(1-f\right)\sin^2{\theta}.
\end{equation}
\end{subequations}

Using these components in Eq. (\ref{deq}) and linear transformation from Eq. (\ref{lt}), we find the equations for tidal forces in free-fall reference frames:
\begin{equation}\label{rd}
    \ddot{\xi}^{r}=-\frac{f''}{2}\xi^{r}=\left(\frac{2M}{r^3}+\frac{\Lambda}{3}\right)\xi^{r},
\end{equation}
\begin{equation}\label{td}
    \ddot{\xi}^{\theta}=-\frac{f'}{2r}\xi^{\theta}=\left(-\frac{M}{r^3}+\frac{\Lambda}{3}\right)\xi^{\theta},
\end{equation}
\begin{equation}\label{pd}
    \ddot{\xi}^{\varphi}=-\frac{f'}{2r}\xi^{\varphi}=\left(-\frac{M}{r^3}+\frac{\Lambda}{3}\right)\xi^{\varphi}.
\end{equation}

This result for $\Lambda = 0$ corresponds to the Schwarzschild case presented in \cite{GR}. It is worth noting that Eq. (\ref{deq}) have four components but we have given expressions only for three: $\xi^{r}$, $\xi^{\theta}$, $\xi^{\varphi}$. The reason for this is that the spectrum of $R^{\mu}_{\nu\alpha\beta}u^{\nu}u^{\alpha}$ has one zero eigenvalue, which gives the insignificant equation $\ddot{\xi}^{t}=0$. We see that the tidal forces in this spacetime depend on the mass of a black hole and value of cosmological constant. We also see that components of tidal force may vanish, in contrast to what happens in the Schwarzschild spacetime $\left(\Lambda=0\right)$. Below we consider the Eqs. (\ref{rd}), (\ref{td}), (\ref{pd}) in more detail.

Eq. (\ref{rd}) gives us zero radial component of tidal force at $r=R^r_0$:
\begin{eqnarray}\label{zrtf}
    R^{r}_0  =  -\sqrt[3]{\frac{6M}{\Lambda}}, \text{or in terms of ($\lambda$, $\rho$)}:\rho_0^{r}  = -\sqrt[3]{\frac{6}{\lambda}},
\end{eqnarray}
which is positive when $\lambda<0$. This means that for a negative cosmological constant, the test body inside the sphere with a radius $R^{r}_0$ undergoes tidal stretch, and in the outside of the sphere with a radius $R^{r}_0$ it experiences tidal compression along the radial direction. For a positive cosmological constant, the test body is subjected only to radial tidal stretch.

From Eq. (\ref{td}) and (\ref{pd}) we get zero of angular components of tidal force at $r=R^a_0$:
\begin{equation}\label{zatf}
    R^{a}_0=\sqrt[3]{\frac{3M}{\Lambda}}, \text{or in terms of ($\lambda$, $\rho$)}:\rho_0^{a}=\sqrt[3]{\frac{3}{\lambda}},
\end{equation}
which is positive when $\lambda>0$. This means that for a positive cosmological constant, the test body inside the sphere with a radius $R^{a}_0$ undergoes tidal angular (transverse) compression, and outside the sphere with a radius $R^{a}_0$ it experiences tidal stretch along the transverse direction. For a negative cosmological constant, the test body is subjected only to angular (transverse) tidal compression.

It is interesting to compare our results for Kottler metric with results for Reissner -- Nordstr{\"{o}}m metric because both these spacetimes are the simplest generalization of Schwarz\-schild metric and have spherical symmetry. The presence of the Reissner -- Nordstr{\"{o}}m black hole charge causes all components of the tidal forces to change sign as well in Kottler case. Unlike Reissner -- Nordstr{\"{o}}m spacetime \cite{TFrn}, all components of tidal force in Kottler metric have no extremes and do not change their monotonicity.
\begin{figure}[h!]
\center{\includegraphics[width = 9.5 cm]{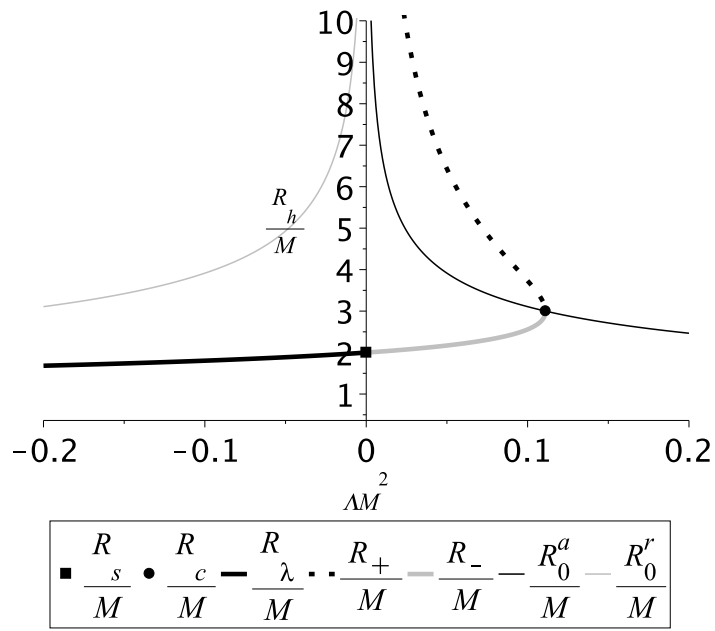}
\caption{\footnotesize{Comparison of the equilibrium radii $\rho_0^a$, $\rho_0^r$ and horizons of the Kottler black hole.}}
\label{bal}}
\end{figure}

In Fig. \ref{bal} we can see $R^r_0>R_{\lambda}$ for all negative values of the cosmological constant, and for value range of cosmological constant where two horizons exist: $R_{-}<R^a_0<R_{+}$.
\subsection{Radial tidal force}
\begin{figure}[h!]
\center{\includegraphics[width = 12 cm]{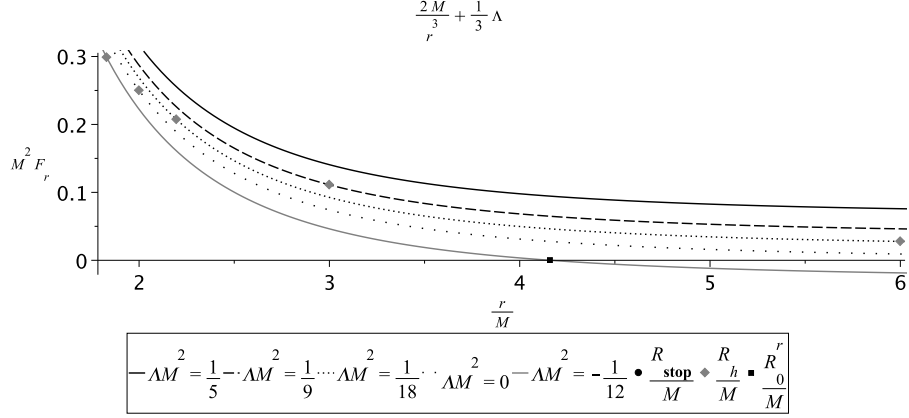}
\caption{\footnotesize{Radial tidal force for Kottler BH with different choices of $\lambda$. We have chosen $b=10M$.}}
\label{radial}}
\end{figure}

In Fig. \ref{radial} we plot the radial tidal force given by Eq. (\ref{rd}) for Kottler black holes for different values of the cosmological constant. The presence of a nonzero $\Lambda$-term does not change the monotonicity of the dependence of tidal force on $\rho=\frac{r}{M}$, it changes only the asymptotic behavior at infinitely large distances. The radial component of the tidal force does not vanish at infinity. At short distances, tidal forces increase as in the case of the Schwarchild black hole ($\Lambda=0$). It illustrates that the radial tidal force changes (for $\Lambda<0$) sign always outside the event horizon, such $R^r_{0}>R_h$.

\begin{figure}[h!]
\center{\includegraphics[width = 12 cm]{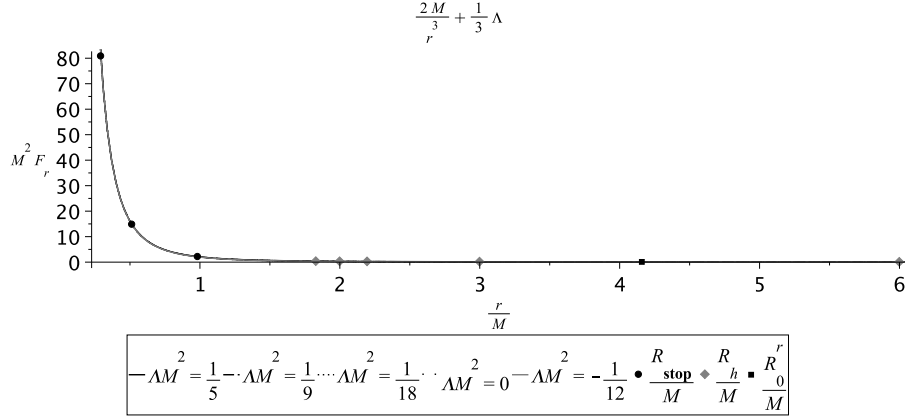}
\caption{\footnotesize{Radial tidal force for Kottler BH with different choices of $\lambda$ on a small scale. We have chosen $b=10M$.}}
\label{radialls}}
\end{figure}

The scale of Fig. \ref{radial} does not allow us to see $R_{stop}$ points on it. Therefore in Fig. \ref{radialls} the same force is shown but on a different scale. And for any positive value of $\lambda$ the radial component of the tidal force is positive for any $r$. Also all points $R_{stop}$, where the radial component of the velocity of a freely falling body is zero, are below the event horizon $R_{stop}<R_h$.

But this scale makes all curves on Fig. \ref{radialls} merge into one.
\subsection{Angular tidal force}
\begin{figure}[h!]
\center{\includegraphics[width = 12 cm]{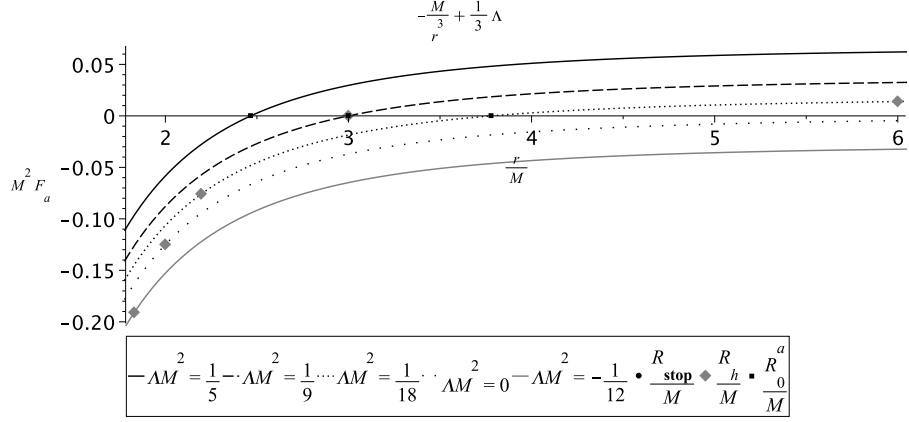}
\caption{\footnotesize{Angular tidal forces for Kottler BH with different choices of $\lambda$. We have chosen $b=10M$.}}
\label{angular}}
\end{figure}

In Fig. \ref{angular} we plot the angular tidal force given by Eq. (\ref{td}) or Eq. (\ref{pd}) (in a spherically symmetric problem azimuthal and polar angles are equal) for Kottler black holes for different values of the cosmological constant. In this graph, we see that the angular component of the tidal force changes sign on the event horizon at a critical value of the cosmological constant $\lambda=\frac{1}{9}$. And for any $\lambda\in\left(0,\frac{1}{9}\right)$ the tidal force changes sign between the two black hole horizons. But at any negative values of $\lambda$, this component of the tidal force does not change sign.

\begin{figure}[h!]
\center{\includegraphics[width = 12 cm]{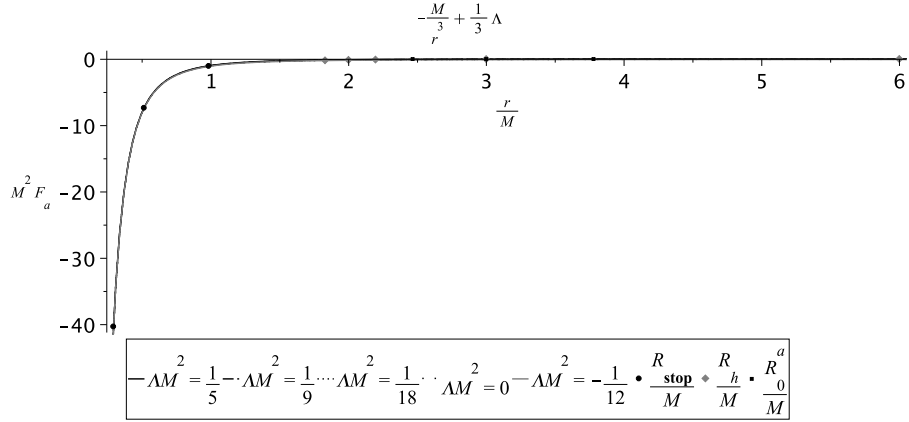}
\caption{\footnotesize{Angular tidal forces for Kottler BH with different choices of $\lambda$ on a small scale. We have chosen $b=10M$.}}
\label{angularls}}
\end{figure}

In Fig. \ref{angularls} we show the same force but on a different scale. These curves demonstrate that the stopping points $R_{stop}$ are inside the event horizon, as well as for the radial component. This is another common property of the Kottler and Reissner -- Nordstr{\"{o}}m metrics. But this scale makes all curves on Fig. \ref{angularls} merge into one.

\section{\label{sec:level5}Solutions of the Geodesic Deviation Equations in Kottler Spacetime}
In this section we solve deviation equations (\ref{rd}), (\ref{td}) and (\ref{pd}) and find the geodesic deviation vectors for radially free-falling geodesics as functions of $r$. We consider a test body infalling radially towards the Kottler black hole. It is simple to convert Eqs. (\ref{rd}), (\ref{td}) and (\ref{pd}) to differential equations in $r$ by using Eq. (\ref{req}). Thus, we find
\begin{equation}\label{rdr}
    \left(E^2-f\right)\frac{d^2\xi^{r}}{dr^2}-\frac{f'}{2}\frac{d\xi^{r}}{dr}+\frac{f''}{2}{\xi}^{r}=0,
\end{equation}
\begin{equation}\label{rda}
   \left(E^2-f\right)\frac{d^2\xi^{a}}{dr^2}-\frac{f'}{2}\frac{d\xi^{a}}{dr}+\frac{f''}{2r}{\xi}^{a}=0,
\end{equation}
where $a=\theta,\varphi$ means angular variables, because Eqs. (\ref{td}) and (\ref{pd}) have the same structure due to spherical symmetry of the problem and prime means the derivative with respect to $r$. In terms of dimensionless variables $\rho$ and $\lambda$ Eqs. (\ref{rdr}) and (\ref{rda}) look accordingly
\begin{equation}\label{drdr}
    \left(E^2-1+\frac{2}{\rho}+\frac{\lambda}{3}\rho^2\right){\xi^r}''+\left(-\frac{1}{\rho^2}+\frac{\lambda}{3}\rho\right){\xi^r}'-\left(\frac{2}{\rho^3}+\frac{\lambda}{3}\right){\xi^r}=0,
\end{equation}
\begin{equation}\label{drda}
   \left(E^2-1+\frac{2}{\rho}+\frac{\lambda}{3}\rho^2\right){\xi^a}''+\left(-\frac{1}{\rho^2}+\frac{\lambda}{3}\rho\right){\xi^a}'-\left(-\frac{1}{\rho^3}+\frac{\lambda}{3}\right){\xi^a}=0,
\end{equation}
where prime means the derivative with respect to dimensionless $\rho$.
The analytical solutions of Eqs. (\ref{drdr}) and (\ref{drda}) can be expressed as quadratures. For the radial component we have
\begin{equation}\label{srdr}
    \xi^{r}(\rho)=\sqrt{E^2-1+\frac{2}{\rho}+\frac{{\lambda}}{3}\rho^2}\left[A+B\int\frac{d\rho}{\left(E^2-1+\frac{2}{\rho}+\frac{{\lambda}}{3}\rho^2\right)^\frac{3}{2}}\right]  ,
\end{equation}
and for the angular components we have
\begin{equation}\label{srda}
    \xi^{a}(\rho)=\rho\left[C+D\int\frac{d\rho}{\rho^2\sqrt{E^2-1+\frac{2}{\rho}+\frac{{\lambda}}{3}\rho^2}}\right],
\end{equation}
where $A$, $B$, $C$ and $D$ are constants of integration. However, the integrals in Eqs. (\ref{srdr}) and (\ref{srda}) are elliptic, which significantly complicates the analysis of these functions.

Fig. 6 and  Fig. 7 show the numerical solutions of differential equations (\ref{drdr}) and (\ref{drda}), respectively. They present the dependence of the component of the geodesic deviation vector on the dimensionless radial variable $\rho$ for different values of $\lambda$.\\

\begin{figure}[h!]
\center{\includegraphics[width = 12 cm]{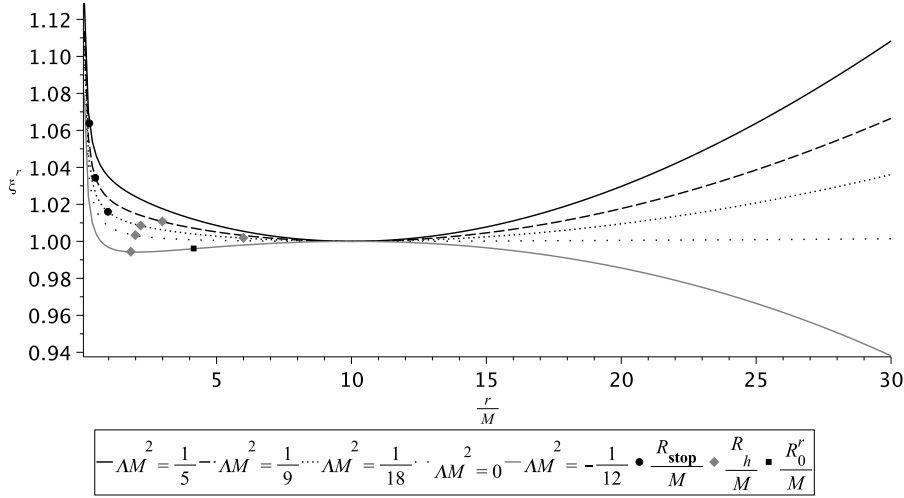}
\caption{\footnotesize{Radial components of the geodesic deviation vector for several values of $\lambda$. We have chosen $b=10M$ and $E=10$, initial conditions: $\xi^{r}(b)=1$, $\xi^{r}{}'(b)=0$.}}
\label{rcvd}}
\end{figure}

The behavior of the radial component of the deviation vector in the Kottler spacetime in the vicinity of the physical singularity is similar to the dependence of the radial component of the deviation vector in the  ($\Lambda=0$). It aims at infinity for $\rho\to 0$. However, far from the center of black hole, the influence of the cosmological constant begins to dominate, therefore, infinitely far from the black hole, the radial geodesic deviation in the Kottler metric differs significantly from asymptotically flat Schwarzschild and Reissner -- Nordstr{\"{o}}m \cite{TFrn} metrics. For $\Lambda>0$ $\xi^r$ grows linearly at $\rho\to\infty$. Case of negative $\lambda$ requires a separate study because the denominator of the integrand from (\ref{srdr}) vanishes.\\

\begin{figure}[h!]
\center{\includegraphics[width = 12 cm]{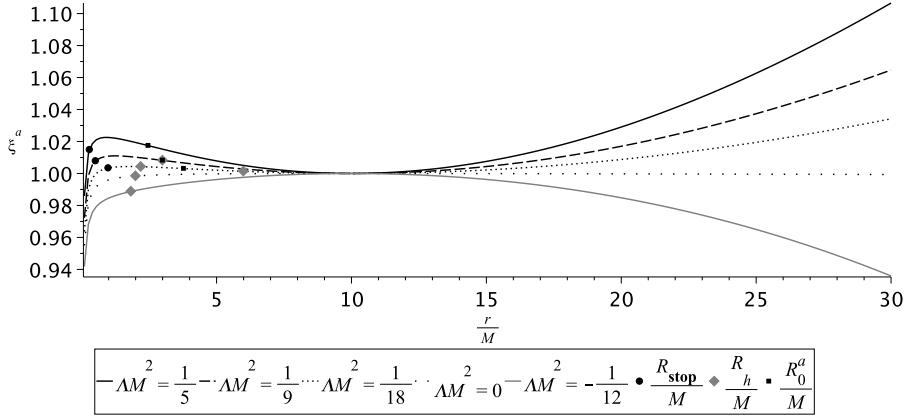}
\caption{\footnotesize{Angular components of the geodesic deviation vector for several values of $\lambda$. We have chosen $b=10M$ and $E=10$, initial conditions: $\xi^{a}(b)=1$, $\xi^{a}{}'(b)=0$.}}
\label{acvd}}
\end{figure}

Angular components $\xi^a$, where $a=\theta,\varphi$, of deviation vector behave like a radial component at spatial infinity. In the vicinity of the center of the black hole they are finite for all values $\lambda$ in contrast to the radial component that is similar to case of Reissner -- Nordstr{\"{o}}m spacetime \cite{TFrn}.
\section{\label{sec:level6}Behavior of solutions near singularity and infinity}
In this section, we give a local description of solutions to the Eqs. (\ref{drdr}) and (\ref{drda}). These are linear homogeneous differential equations with variable coefficients. Since we have elliptic integrals in Eqs. (\ref{srdr}) and (\ref{srda}), so we would like to describe the local behavior of solutions in the surrounding area of the singularity of a black hole and infinitely far from it. We show how expressions (\ref{srdr}) and (\ref{srda}) behave for $\rho\to 0$ and $\rho\to \infty$.
\subsection{Radial component}
To determine the behavior of $\xi^{r}$ for small or large $\rho$ let us consider Taylor series at $\rho\to 0$ and $\rho \to \infty$ (or $\frac{1}{\rho}\to 0$) respectively. In Eq. (\ref{srdr}), the root and the integrand are expanded in a series
\begin{subequations}\label{ras}
\begin{multline}\label{rass}
 \xi^{r}(\rho)=\left(\sqrt{\frac{2}{\rho}}+\frac{\sqrt{2}}{4}\left(E^2-1\right)\sqrt{\rho}+O\left(\rho^{\frac{3}{2}}\right)\right)\times \\ \times
 \left[A+B\int \left\{\frac{\sqrt{2}}{4}\rho^{\frac{3}{2}}-\frac{3\sqrt{2}}{16}\left(E^2-1\right)\rho^{\frac{5}{2}}+O\left(\rho^{\frac{7}{2}}\right)\right\}d\rho\right]\;\text{for $\rho\to0$},
\end{multline}
\begin{multline}\label{rasl}
\xi^{r}\left(\rho\right)=\left(\sqrt{\frac{\lambda}{3}}\rho+\sqrt{\frac{3}{\lambda}}\frac{E^2-1}{2\rho}+O\left(\frac{1}{\rho^2}\right)\right)\times \\ \times\left[A+\left(\frac{3}{\lambda}\right)^{\frac{3}{2}}B\int \left\{\frac{1}{\rho^3}+\frac{3}{\lambda}\frac{E^2-1}{2\rho^5}+O\left(\frac{1}{\rho^6}\right)\right\}d\rho\right]\;\text{for $\rho\to\infty$}.
\end{multline}
\end{subequations}
After integrating the sum of power functions and multiplying brackets in both expressions (\ref{rass}) and (\ref{rasl}) we get
\begin{subequations}\label{sras}
\begin{equation}\label{srass}
 \xi^{r}(\rho)=\frac{\sqrt{2}A}{\sqrt{\rho}}+\frac{\sqrt{2}A\left(E^2-1\right)}{4}\sqrt{\rho}+O\left(\rho^{\frac{3}{2}}\right)\;\text{for $\rho\to 0$},
 \end{equation}
\begin{equation}\label{srasl}
\xi^{r}\left(\rho\right)=\sqrt{\frac{\lambda}{3}}A\rho+\left[\sqrt{\frac{3}{\lambda}}\frac{A\left(E^2-1\right)}{2}-\frac{3B}{2\lambda}\right]\frac{1}{\rho}+O\left(\frac{1}{\rho^2}\right)\;\text{for $\rho\to \infty$}.
\end{equation}
\end{subequations}
From Eq. (\ref{srass}) we see that near the black hole singularity the radial component of deviation vector tends to infinity as $\frac{1}{\sqrt{\rho}}$, whereas according to Eq. (\ref{srasl}) at spatial infinity it grows linearly only for positive $\lambda$. The asymptotic behavior of $\xi^r(\rho)$ from (\ref{srass}) and (\ref{srasl}) completely coincides with the numerical solution shown in Fig. \ref{rcvd}, except for the case when $\lambda<0$. Numerical solutions from Fig. \ref{rcvd} correspond to (\ref{srass}) and (\ref{srasl}) in the indicated areas of $\rho$.
\subsection{Angular components}
A similar analysis of Eq. (\ref{srda}) in the same way leads first to
\begin{subequations}
\begin{equation}
\xi^{a}(\rho)=\rho\left[C+D\int\left(\frac{\rho^{-\frac{3}{2}}}{\sqrt{2}}-\frac{E^2-1}{4\sqrt{2}}\rho^{-\frac{1}{2}}+O\left(\sqrt{\rho}\right)\right)d\rho\right]\; \text{for $\rho\to0$},
\end{equation}
\begin{equation}
\xi^{a}(\rho)=\rho C+\rho D\sqrt{\frac{3}{\lambda}}\int\left\{\frac{1}{\rho^3}-\frac{3}{\lambda}\frac{E^2-1}{2\rho^5}+O\left(\frac{1}{\rho^6}\right)\right\}d\rho\;\text{for $\rho\to\infty$},
\end{equation}
\end{subequations}
and after integration and multiplication to
\begin{subequations}\label{aas}
\begin{equation}\label{aass}
\xi^{a}(\rho)=-\sqrt{2}D\sqrt{\rho}+C\rho+O\left(\rho^{\frac{3}{2}}\right)\;\text{for $\rho\to0$},
\end{equation}
\begin{equation}\label{aasl}
\xi^{a}(\rho)=C\rho-\sqrt{\frac{3}{\lambda}}\frac{D}{2\rho}+O\left(\frac{1}{\rho^3}\right)\;\text{for $\rho\to\infty$}.
\end{equation}
\end{subequations}
Eq. (\ref{aass}) means the angular components of the deviation vector near the black hole singularity grows as $\sqrt{\rho}$ (this is different from Eq. (\ref{srass})), while the linear growth at infinity in Eq. (\ref{aasl}) is similar Eq. (\ref{srasl}), despite the absence of a $\lambda$ leading order. The asymptotic behavior of $\xi^a(\rho)$ from (\ref{aass}) and (\ref{aasl}) completely coincides with the numerical solution shown in Fig. \ref{acvd}, except for the case when $\lambda<0$. Numerical solutions from Fig. \ref{acvd} correspond to (\ref{aass}) and (\ref{aasl}) in the indicated areas of $\rho$.
\section{Separately about Anti-de Sitter}
\begin{figure}[h!]
\center{\includegraphics[width = 12 cm]{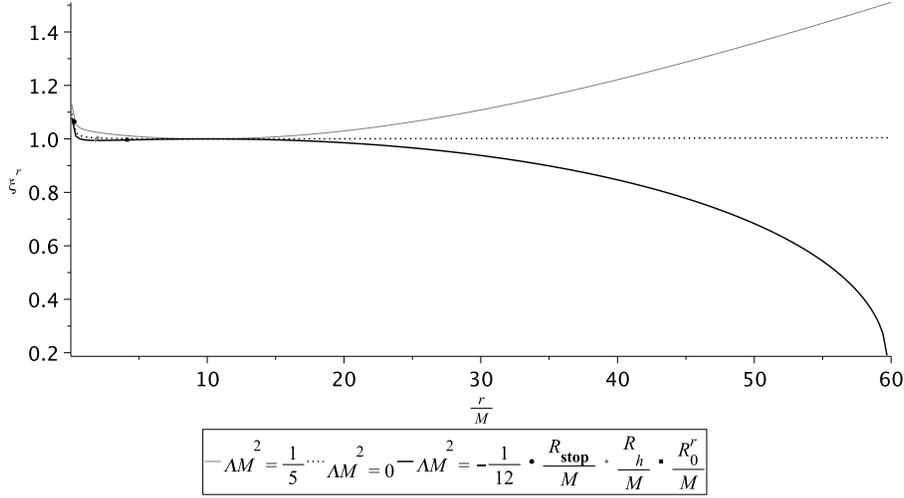}
\caption{\footnotesize{Radial components of the geodesic deviation vector for several values of $\lambda$. We have chosen $b=10M$ and $E=10$, initial conditions: $\xi^{r}(b)=1$, $\xi^{r}{}'(b)=0$.}}
\label{rsing}}
\end{figure}

\begin{figure}[h!]
\center{\includegraphics[width = 12 cm]{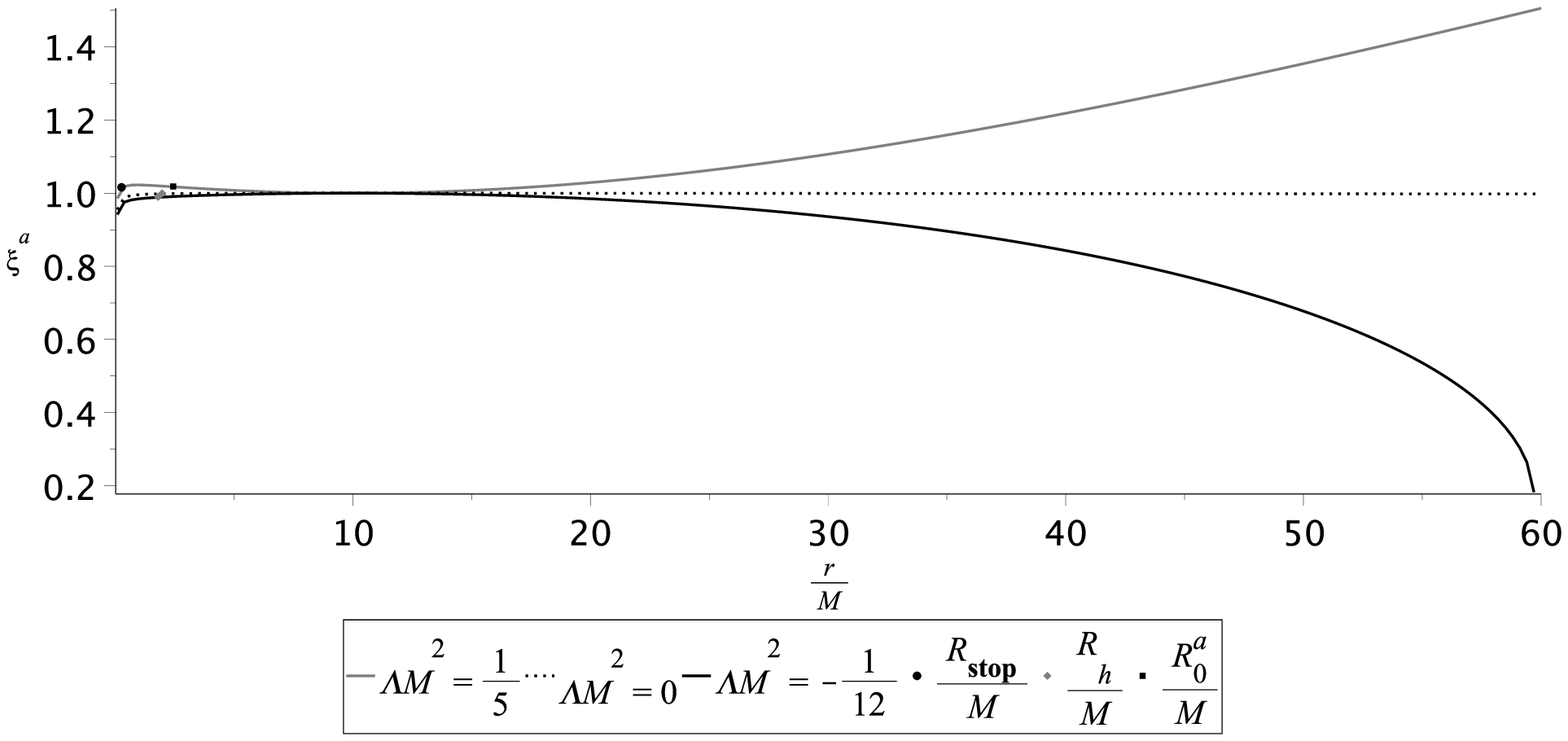}
\caption{\footnotesize{Angular components of the geodesic deviation vector for several values of $\lambda$. We have chosen $b=10M$ and $E=10$, initial conditions: $\xi^{a}(b)=1$, $\xi^{a}{}'(b)=0$.}}
\label{asing}}
\end{figure}

Now let us consider the Anti-de Sitter spacetime where $\lambda<0$ ($\lambda=-|\lambda|$). In this case we can see that the denominator of expressions (\ref{srdr}) and (\ref{srda}) can vanish.
\begin{equation}\label{eqsing}
E^2-1+\frac{2}{\rho}-\frac{{|\lambda|}}{3}\rho^2=0.
\end{equation}
Eq. (\ref{eqsing}) reduces to the cubic equation
\begin{equation}\label{cubsing}
\rho^3-\frac{3(E^2-1)}{|\lambda|}\rho-\frac{6}{|\lambda|}=0.
\end{equation}
Using the Cardano formula we get
\begin{equation}\label{cubroot}
\rho_0=\sqrt[3]{\frac{3}{|\lambda|}+\frac{3}{|\lambda|}\sqrt{1-\frac{(E^2-1)^3}{9|\lambda|}}}+\sqrt[3]{\frac{3}{|\lambda|}-\frac{3}{|\lambda|}\sqrt{1-\frac{(E^2-1)^3}{9|\lambda|}}}.
\end{equation}
The nature of the roots of this equation depends on the sign of the expression $1-\frac{(E^2-1)^3}{9|\lambda|}$:
\begin{enumerate}
\item $1>\frac{(E^2-1)^3}{9|\lambda|}$ means that Eq. (\ref{cubsing}) has a single real positive simple root $\rho_0$ described by the Eq. (\ref{cubroot}).
\item $1=\frac{(E^2-1)^3}{9|\lambda|}$ means that Eq. (\ref{cubsing}) takes the form
\begin{equation}\label{crcubeq}
\rho^3-\frac{27}{(E^2-1)^2}\rho-\frac{54}{(E^2-1)^3}=\left(\rho-\frac{6}{E^2-1}\right)\left(\rho+\frac{3}{E^2-1}\right)^2=0,
\end{equation}
which has only one real simple positive root $\rho=\frac{6}{E^2-1}$, but the second root $\rho=-\frac{3}{E^2-1}$ is negative what is pointless for us.
\item $1<\frac{(E^2-1)^3}{9|\lambda|}$ means that Eq. (\ref{cubsing}) has three real root
\begin{equation}
\rho=2\sqrt{\frac{E^2-1}{|\lambda|}}\cos\left(\frac{\arccos\left(\frac{3\sqrt{|\lambda|}}{(E^2-1)^\frac{3}{2}}\right)+2\pi k}{3}\right),
\end{equation}
where $k=0,1,2$. And among them, only one root is positive. This is the root for $k=0$.
\end{enumerate}
Thus, all the cases described above show us that for any $\lambda<0$ there is simple positive root of Eq. (\ref{eqsing}). Let us label this single simple root $\rho_0$, so the left side of the Eq. (\ref{eqsing}) can be represented as
\begin{equation}\label{locsqr}
E^2-1+\frac{2}{\rho}-\frac{{|\lambda|}}{3}\rho^2\equiv(\rho_0-\rho)Q(\rho).
\end{equation}
Where $Q(\rho)$ is rational function which has no positive roots. Expression (\ref{locsqr}) allows us to explore (\ref{srdr}) and (\ref{srda}) in the vicinity of $\rho_0$. Representing square roots and integrands as a Taylor series, we can, after integration, obtain the local behavior of the deviation vector components $\xi^r(\rho)$ and $\xi^a(\rho)$
\begin{equation}\label{locdr}
    \xi^{r}(\rho)=\frac{2B}{Q(\rho_0)}+A\sqrt{(\rho_0-\rho)Q(\rho_0)}+O(\rho_0-\rho)\;\text{for $\rho\to\rho_0-0$},
\end{equation}
\begin{equation}\label{locda}
    \xi^{a}(\rho)=\rho_0C-\frac{2D\sqrt{\rho_0-\rho}}{\rho_0\sqrt{Q(\rho_0)}}+O\left(\rho_0-\rho\right)\;\text{for $\rho\to\rho_0-0$}.
\end{equation}
Thus, we found a root feature for all components of the deviation vector. If we extend the numerical solution of Eqs. (\ref{drdr}) and (\ref{drda}) shown in Fig. \ref{rcvd} and Fig. \ref{acvd} to longer distances we can demonstrate in Fig. \ref{rsing} and Fig. \ref{asing} that for negative $\lambda$ $\xi^{r}$ and $\xi^{a}$ behave in a root manner.


Thus Eqs. (\ref{locdr}) and (\ref{locda}) accurately describe the behavior of the bottom bold curves in Fig. \ref{rsing} and Fig. \ref{asing} in the vicinity of the singular points $\rho=\rho_0$. The reason for non-analytical behavior of the components of the geodesic deviation vector is that geodesic equation (\ref{teq}) for $\lambda<0$ defined only for $\rho\le\rho_0$. It means that negative energy density leads to tidal compression along all spatial directions at large distances from the black hole.
\section{Conclusion}
In this article we discuss the Kottler black hole which is solution of Einstein's equations when the cosmological constant plays the role of a matter. We analyze properties of Kottler spacetime and build the dependence of event horisons radius on cosmological constant value and describe properties of radial geodesics in Kottler metric. We detected the presence of stopping points in the process of free movement in the investigated space (\ref{rstop}). Using geodesic deviation equation we study tidal force in Kottler black hole, we have shown that they can change sign (\ref{zrtf}) and (\ref{zatf}) as in the Reissner -- Nordstr{\"{o}}m spacetime in contrast to the Schwarzschild metric. It was shown that unlike Schwarzschild spacetime ($\Lambda$ = 0), in the case of Kottler metric both radial and angular components of tidal force may vanish.  But tidal force in the Kottler and Schwarzschild spaces also have common features different from charged Reissner -- Nordstr{\"{o}}m black hole, they do not have an extrema and therefore do not change their monotonicity. The nonzero $\Lambda$-term changes the asymptotic behavior of tidal force at infinitely large distances (\ref{rd}), (\ref{td}) and (\ref{pd}): the radial and angular components do not vanish at infinity. However, at short distances tidal forces increase as in the case of the Schwarzschild black hole. The analytical solutions of the geodesic deviation equations expressed as quadratures were founded (\ref{srdr}) and (\ref{srda}). For these analytical solutions, we managed to obtain expressions describing the asymptotic behavior of the deviation vector components near the singularity of the black hole (\ref{srass}), (\ref{aass}) and at large distances from it (\ref{srasl}), (\ref{aasl}). The solutions of the geodesic deviation equations in the Schwarzschild -- Anti-de Sitter spacetime (Kottler with $\Lambda<0$) are considered separately. With a negative value of the cosmological constant, it was possible to detect a root peculiarity in the behavior of the deviation vector at large distances (\ref{locdr}) and (\ref{locda}) in comparison with the radius of the event horizon. This means that, in Schwarzschild -- Anti-de Sitter spacetime, there is a region of tidal compression at a great distance from the black hole. Analytical calculations are confirmed by the numerical solution of the equations for the deviation of geodesics.

Development of tidal effects studying in general relativity can have several directions. First of all, it is interesting to consider any combinations of matter, for example, quintessence and cosmological constant together. Secondary it can be investigated how tidal forces are affected by the presence of non-zero angular momentum of the freely falling body. The third direction in the study of tidal forces near black holes can be the study of the properties of geodesic deviation equations in multidimensional spacetime. It is also very interesting to observe how tidal forces behave in axially symmetric spaces in the presence of matter. A separate completely unexplored question is the nature of tidal forces in the metrics of black rings. As we can see, despite the fact that the classical effects in the general theory of relativity have been studied for more than a hundred years, this issue still contains a large number of unsolved problems.

\end{document}